\RequirePackage{fix-cm}
\documentclass[smallextended]{svjour3}       % onecolumn (second format)
\smartqed  % flush right qed marks, e.g. at end of proof
\expandafter\let\csname equation*\endcsname=\relax
\expandafter\let\csname endequation*\endcsname=\relaxD
\usepackage{epsf,graphicx,graphics}
\usepackage{amsmath}
\usepackage{latexsym,amssymb}
\usepackage{setspace,cite}
\usepackage{a4}
\usepackage{slashed}
\usepackage{morefloats}
\usepackage{float}
\usepackage[normal]{caption}

\usepackage[left=4cm,right=4cm,top=2cm,bottom=3cm]{geometry}

\begin{document}

%Makes theorem, prop etc number correctly
\newtheorem{thr}{Theorem}
\newtheorem{defn}[thr]{Definition}
\newtheorem{lem}[thr]{Lemma}
\newtheorem{prop}[thr]{Proposition}
\newtheorem{cor}[thr]{Corollary}

%\disfrac makes fractions look nice if they display too small
\newcommand{\disfrac}[2]{\displaystyle{\frac{#1}{#2}}}

%Connection coefficients Gamma
\newcommand{\con}[3]{\Gamma^{#1}_{#2 #3}}

%Lie algebra shorthand
\newcommand{\uone}{\mathfrak{u}(1)}
\newcommand{\essu}{\mathfrak{su}(2)}
\newcommand{\sun}{\mathfrak{su}(N)}

%bold & underlined vectors
\newcommand{\bv}[1]{\underline{\mbox{\boldmath$#1$}}}

%r-star partial derivs
\newcommand{\drs}{\partial_{r_*}}
\newcommand{\drss}{\partial_{r_*}^2}

%The set of reals
\newcommand{\R}{\mathbb{R}}

\newcommand{\newpar}{\\ \\}

\renewcommand{\thefootnote}{\arabic{footnote}}

\title{On the existence of topological hairy black holes in $\sun$ EYM theory with a negative cosmological constant%\thanks{Grants or other notes
}
%\subtitle{Published in Gen.Relativ.Gravit. (2015) 47:1829;\\ DOI 10.1007/s10714-014-1829-5}

\titlerunning{Topological $\sun$ hairy black holes with $\Lambda<0$}        % if too long for running head

\author{J. Erik Baxter         %\and
        %Second Author %etc.
}


\institute{J. Erik Baxter \at
              Harmer Building, \\
			  Sheffield Hallam University, \\
			  Pond Street, \\
			  Sheffield, \\
			  South Yorkshire \\ 
			  S1 1WB \\
              \email{e.baxter@shu.ac.uk}           %  \\
           %\and
           %S. Author \at
            %  second address
}

\date{Published in Gen.Relativ.Gravit. (2015) 47:1829\\ DOI 10.1007/s10714-014-1829-5\\Received: 28th August 2014 / Accepted: 23rd October 2014}

\maketitle

\begin{abstract}
We investigate the existence of black hole solutions of four dimensional $\mathfrak{su}(N)$ EYM theory with a negative cosmological constant. Our analysis differs from previous works in that we generalise the field equations to certain non-spherically symmetric spacetimes. The work can be divided into two sections. In the first half, we use theorems of Wang's to derive a new $\sun$-invariant one-form connection which is appropriate for certain non-spherically symmetric spacetimes; this will serve as the new ansatz for our gauge potential. The second half is devoted to proving the existence of non-trivial solutions to the field equations for any integer $N$, with $N-1$ gauge degrees of freedom. Specifically, we prove existence in two separate regimes: for fixed values of the initial parameters and as $|\Lambda|\rightarrow\infty$; and for any $\Lambda<0$, in some neighbourhood of existing trivial solutions. In both cases we can prove the existence of `nodeless' solutions, i.e. such that all gauge field functions have no zeroes; this fact is of interest as we anticipate that some of them may be stable.
\end{abstract}

\section{Introduction}

Until relatively recently, the question of classifying black holes seemed resolved, with Birkhoff's theorem \cite{47} and uniqueness theorems of Israel's for Schwarzschild \cite{63} and Reissner-N\"{o}rdstrom \cite{65} black holes. This lead to Wheeler's so-called `no-hair' conjecture \cite{48}, which placed limits of the kinds of black holes one could expect in spherically symmetric space (and specifically the very small number of parameters which must entirely specify the solution). However, in the 1980s and since, more and more examples have been found which show that the conjecture is violated in the strictest terms, beginning with the discovery of black hole and soliton solutions in 4D flat spacetime; in the case of scalar hair by Bekenstein and others (e.g. \cite{57, 58, 59}), and for $\mathfrak{su}(2)$ hair by Bizon, Bartnik, McKinnon and others (e.g. \cite{3, 4, 61, 62}). Since then, many other directions of research in this area have been pursued (see for instance \cite{5} for a review.)

Introducing a non-zero cosmological constant is known to yield very interesting results, due to the differing geometry it bestows upon the manifold. In asymptotically flat space, soliton and black hole solutions of $\essu$ EYM theory are known to exist only at discrete points in the parameter space. All such solutions will have a gauge field function $\omega$ with at least one zero, the number of unstable modes of the solution being directly proportional to the number of zeroes of $\omega$ ($2k$ unstable modes for $k$ zeroes \cite{15, 45, 46}). However, for anti-de Sitter (adS) space (i.e. a negative cosmological constant), it changes completely; we get solutions existing in continuous ranges of the parameters governing the dynamic variables \cite{12, 13, 14, 38}, and at least some of them possess gauge fields with no zeroes (so-called `nodeless' solutions). It can be shown that some of the solutions with non-zero gauge fields everywhere are stable under linear perturbations \cite{12, 13, 14, 16, 17, 10}, provided that $|\Lambda|$ is large enough. (We shall leave the question of $\Lambda>0$ entirely, except to say that for a comprehensive classification of constructed asymptotically de Sitter spacetimes, see \cite{7}.)

Recent research has expanded upon this, and the existence has been established of asymptotically anti-de Sitter soliton and black hole solutions to 4D $\mathfrak{su}(N)$ EYM equations \cite{18, 11, 19}. Purely magnetic solutions have been found which are described by $N-1$ gauge field functions $\omega_j$ ($j=1,...,N-1$). The most recent results we have \cite{1} show that there exist genuinely non-trivial solutions for any $N$ where the gauge field functions have no zeroes, and that for $|\Lambda|$ sufficiently large, at least some of these are stable \cite{19, 10}. These hairy black holes in adS, despite being indistinguishable from Reissner-N\"{o}rdstrom black holes asymptotically, therefore require an extra $N-1$ parameters to describe them, and this number is obviously unbounded (though still countable) as $N\rightarrow\infty$. For more detail, see the recent work \cite{49}, in which there do seem to be global charges characterising at least some of the stable $\mathfrak{su}(N)$ solutions.

Some work from van der Bij and Radu \cite{2} (among others) has suggested a whole new direction in which to take these `hairy' black holes. Their work is on topological black holes -- that is, black holes that possess isometries other than the usual requirement of spherical symmetry. This will be expanded upon later, but briefly, the requirement is no longer that the metric and gauge potential are \textit{spherically} symmetric (i.e. endow the manifold with gauge fields that are invariant under an action of $SU(2)$ by principal bundle automorphisms); it is relaxed to encompass other isometry groups such that the Lie group of isometries is the structure group of one of the three surfaces of constant curvature -- i.e., the sphere, the plane, or a hyperbolic surface (with the possibilities of event horizons of non-zero genus). Several authors have also considered various other cases of topological black holes, including those set in higher dimensions and with event horizons of genus $g\geq 1$ (e.g. \cite{8, 23, 50, 51, 52, 53, 54, 55, 66}).

This work \cite{2} yielded interesting results for $\mathfrak{su}(2)$: the authors found topological black hole solutions numerically; noted that the Yang-Mills field equation could be cast in a form in which it was obvious that all solutions would be nodeless; and demonstrated their stability under linear spherically symmetric perturbations for $|\Lambda|$ sufficiently large. It is the aim of this paper to build on their successes (though using largely different methods) by generalising their results to $\mathfrak{su}(N)$; and thus to prove

\begin{enumerate}

\item that, for $|\Lambda|\rightarrow\infty$, solutions exist for arbitrary values of the initial parameters of the field variables at the event horizon which are regular everywhere and which are analytic in their parameters, and

\item that genuine (i.e. non-trivial) solutions to the 4D $\mathfrak{su}(N)$ EYM field equations with a negative cosmological constant can be found in a neighbourhood of known (trivial) solutions and, for $k\neq 1$, these solutions will be nodeless i.e. will possess gauge field functions with no zeroes.

\end{enumerate}

The results for $k=1$ (which correspond to spherical symmetry) have been adequately explored in \cite{1}, and so will largely be ignored except as comparison to the new results we obtain for $k\neq 1$. Also, we note that here we use a time-honoured `shooting' method to verify the existence of solutions, though this is not the only method: recent work by Nolan and Winstanley \cite{50} uses an argument involving Banach spaces to prove the existence of dyons and dyonic black holes in spherically symmetric $\mathfrak{su}(2)$ theory.

The outline of this paper is as follows. First, we present the ans\"{a}tze, field equations and boundary conditions for 4D $\mathfrak{su}(N)$ EYM theory with a negative cosmological constant. We use the work of Wang \cite{31} and K\"{u}nzle \cite{21} to derive a new, more general $\sun$-invariant gauge potential which applies to the topologies we are interested in. Some trivial solutions to the field equations are found, including Reissner-N\"{o}rdstrom-type and embedded $\mathfrak{su}(2)$ solutions, both of which are vital to the final existence argument.

Next we prove a series of propositions: we demonstrate that solutions exist locally nearby the singular points of the field equations (that is, $r=r_h$ and $r\rightarrow\infty$); and that solutions which are regular in some small interval within $r_h<r<\infty$ can continue to be integrated outwards to the asymptotic regime provided the metric function $\mu>0$. We note that such solutions will be analytic in their initial values at the event horizon. We then use these propositions to prove a theorem which states that topological $\mathfrak{su}(N)$ solutions will always exist in open sets, and that solutions in a sufficiently small neighbourhood of each other will possess the same number of gauge field zeroes.

Finally we establish our two main results: that non-trivial solutions to these field equations exist both for fixed values of the boundary conditions $\omega_{j,h}$ and for $|\Lambda|\rightarrow\infty$, and also in some small neighbourhood of Reissner-N\"{o}rdstrom-type solutions for any fixed $\Lambda<0$. The conclusions are presented at the end along with an indication of future directions this work may take.

\section{2D spaces of constant curvature}\label{2dcc}

In this section we give a brief review of the three different topological cases of 2-dimensional spaces of constant Gaussian curvature $K$ (i.e. spherical, planar and hyperboloidal), including a discussion of the metrics and the Lie groups which describe the symmetries of the spaces. 

\subsection{Metrics and isometry groups: 3 different cases}

We begin with a manifold $M$ endowed with a metric, possessing some symmetries of its own; and wish to endow it with fields possessing some other symmetry group $G$. We can consider it as a principal fibre bundle, i.e. with base space $M$ and $G$-valued fibres; and endow that bundle with a connection which possesses the property that it is invariant under any isometric action (automorphism) performed on it. In the case of spherical symmetry, such an action would be a rotation around a general axis. It can be noted that each such action can be mapped to the (continuous) rotation group $U(1)$ -- as such, this is associated with the $\mathfrak{u}(1)$ subalgebra. In the case of a sphere, the collection of all these actions form orbit surfaces which in the trivial case are clearly spheres $S^2$, which regularly foliate the 4-manifold. The whole group of isometries is isomorphic to $SU(2)$. 

In the case of a general space of constant curvature, we find that the $U(1)$ subgroup is still present in $LG(2)$ (the general name we are giving to the Lie group of isometries of the 2D constant-curvature manifold in question), and since the collection of all possible actions of $LG(2)$ regularly foliates the 4-manifold in question, then the isometries are classified by all possible actions of $LG(2)$ on a principal $G$-bundle. In order to generalise the above statements, and the arguments that follow it in the original paper by K\"{u}nzle \cite{21}, to $k\neq 1$, we will need appropriate replacements for the subspace $S^2$ and the Lie group of isometries $SU(2)$. Then we will have a way forward when it comes to constructing the gauge potential, in order to give the gauge fields the correct symmetry. Hence, in what shortly follows we shall briefly explain what we mean by surfaces of constant curvature, categorising them by the sign of the Gaussian curvature, and showing the appropriate topology and Lie algebra we use in each case.

\subsubsection{$k=1$: The sphere}

As is well known, the metric for the sphere of unit radius is given by
\begin{equation}\label{k1met}
d\Omega_1^2 = d\theta^2+\sin^2\theta d\phi^2.
\end{equation}
The symmetry group is also well known, as the Lie group $SU(2)$, with the associated Lie algebra $\mathfrak{su}(2)$\footnote{Henceforth Lie groups will be denoted by Latin script, and their associated algebras by Germanic script.}. The most common representation of this algebra is to use the well-known \textit{Pauli matrices}, i.e.:

\begin{equation}\begin{array}{ccc}
\sigma_1=\left( \begin{array}{cc}
0 & 1 \\
 1 & 0 \end{array}\right) &
\sigma_2=\left(\begin{array}{cc}
0 & -i \\ 
i & 0 \end{array}\right) &
\sigma_3=\left(\begin{array}{cc}
1 & 0 \\ 
0 & -1 \end{array}\right),
\end{array}
\end{equation}
to obtain the three generators. It should be noted that the Lie group $SU(2)$ forms a double cover of the Lie group $SO(3)$; thus, the three generators of the algebra $\mathfrak{su}(2)$ can be mapped to the three (infinitesimal) basis vectors for ordinary 3-dimensional Euclidean spatial rotations. If we let
\begin{equation}
K_j = \frac{-i\sigma_j}{2}
\end{equation}
with $j\in \{1,2,3\}$, then the $K_j$ obey the following commutation relations:
\begin{equation}\label{k1cr}
\begin{array}{ccccc}
[K_1,K_2]=K_3 & \quad &
[K_2,K_3]=K_1 & \quad &
[K_3,K_1]=K_2.
\end{array}
\end{equation}
\par
As well as using the Pauli matrices as a basis, we can use the following definitions of infinitesimal differential operators to satisfy the commutators \eqref{k1cr}:
\begin{equation}
\begin{split}
K_1&\mapsto\frac{\partial}{\partial\alpha}=z\frac{\partial}{\partial y} - y\frac{\partial}{\partial z},\\
K_2&\mapsto\frac{\partial}{\partial\beta}=x\frac{\partial}{\partial z} -z\frac{\partial}{\partial x},\\
K_3&\mapsto\frac{\partial}{\partial\phi}=y\frac{\partial}{\partial x} - x\frac{\partial}{\partial y};
\end{split}
\end{equation}
where $\partial/\partial\alpha$, $\partial/\partial\beta$ and $\partial/\partial\phi$ can be mapped to ordinary $SO(3)$ rotations about the $x$, $y$ and $z$ axes. 

\subsubsection{$k=0$: The plane}

The metric for a 2D plane can be given as the following:
\begin{equation}\label{k0met}
d\Omega_0^2 = d\theta^2+\theta^2d\phi^2.
\end{equation}
The isometry group for the 2D plane is just the ordinary 2D Euclidean Lie group $E(2)$, with associated Lie algebra $\mathfrak{e}(2)$. The Lie group $E(2)$ has three generators: two spatial translations and one rotation. The basis for this is most helpfully taken as infinitesimal translations in the $x$ and the $y$ directions, which for later convenience we call $K_1$ and $K_2$; and the rotation in the $xy$ plane, for which we use the name $K_3$. The commutation relations for $\mathfrak{e}(2)$ are then:
\begin{equation}\label{k0cr}
\begin{array}{ccccc}
[K_1,K_2]=0, & \quad &
[K_2,K_3]=K_1, & \quad &
[K_3,K_1]=K_2.
\end{array}
\end{equation}
Unfortunately, the isometries of $\R^2$ cannot be given explicitly in terms of matrices under multiplication -- the closest we can get is the set of transforms
\begin{equation}\label{ktrans}
\begin{array}{rl}
\left(\begin{array}{c}
x\\
y \end{array}\right)\mapsto & \left(\begin{array}{c}
x+a\\
y\\ \end{array}\right)\\
\left(\begin{array}{c}
x\\
y \end{array}\right)\mapsto & \left(\begin{array}{c}
x\\
y+b\\ \end{array}\right)\\
\left(\begin{array}{c}
x\\
y \end{array}\right)\mapsto & R_\theta\left(\begin{array}{c}
x\\
y \end{array}\right)=\left(\begin{array}{cc}
\sin\theta & \cos\theta\\
\cos\theta & -\sin\theta
\end{array}\right)\left(\begin{array}{c}
x\\
y \end{array}\right)
\end{array}
\end{equation}
for a translation in the plane by the vector 
\begin{equation}
\left(\begin{array}{c}
a\\
b \end{array}\right)
\end{equation}
and a rotation by an angle $\theta$. So the best representation for this algebra can be given in terms of the following correspondence to differential operators:
\begin{equation}\label{kdiff}
\begin{split}
K_1&\mapsto\frac{\partial}{\partial x},\\ 
K_2&\mapsto\frac{\partial}{\partial y},\\
K_3&\mapsto\frac{\partial}{\partial\phi}=y\frac{\partial}{\partial x} - x\frac{\partial}{\partial y}.
\end{split}
\end{equation}
These are just the ordinary killing vectors for 2-dimensional Euclidean space.

\subsubsection{$k=-1$: The hyperboloid}

The analogous negatively curved space is endowed with the metric
\begin{equation}\label{k-met}
d\Omega_{-1}^2 = d\theta^2+\sinh^2\theta d\phi^2,
\end{equation}
where $d\Omega_{-1}^2$ is regarded as the metric on the sphere of unit imaginary radius, i.e. the unit hyperboloid. The symmetry group on the space can be shown to be the Lie group $SU(1,1)$, associated with the Lie algebra $\mathfrak{su}(1,1)$. The Lie algebra $\mathfrak{su}(1,1)$ again has 3 infinitesimal generators which we call $K_j$ for $j\in \{1,2,3\}$; and can be cast in the form where they have the following commutation relations (see e.g. \cite{22, 60}):
\begin{equation}\label{k-cr}
\begin{array}{ccccc}
[K_1,K_2]=-K_3,& \quad &
[K_2,K_3]=K_1,& \quad &
[K_3,K_1]=K_2.
\end{array}
\end{equation}
A representation for this can be given in terms of the following matrices:
\begin{equation}\label{su11map}\begin{array}{c}
K_1\mapsto -\frac{i}{2}\left( \begin{array}{cc}
0 & -i \\
-i & 0 \end{array}\right),\\
K_2\mapsto -\frac{i}{2}\left(\begin{array}{cc}
0 & -1 \\ 
1 & 0 \end{array}\right),\\
K_3\mapsto -\frac{i}{2}\left(\begin{array}{cc}
1 & 0 \\ 
0 & -1 \end{array}\right).
\end{array}
\end{equation}
Note that $K_3$ is identical to the definition of $K_3$ for the spherical case -- this is no coincidence and will be discussed below.

\subsection{Topological considerations}

In all 3 cases, giving the appropriate symmetry ($SU(2)$, $E(2)$ or $SU(1,1)$) to the 4D manifold requires that they be considered regularly foliated by 2D surfaces of constant curvature (which we shall call $\Sigma_k^2$) that are appropriate to the case in question, endowing them with the topology $\mathbb{R}^2\times\Sigma_k^2$. 

In the case of the sphere (for $k=1$), it is clear that $\Sigma_1^2=S^2$, and it can be shown that $S^2\simeq SU(2)/U(1)$. $SU(2)$ as a manifold is topologically the 3-sphere $S^3$, and the quotient is taken using $U(1)$ (which is the circle $S^1$). To form this quotient we can use the fact that the Abelian subgroup of diagonal matrices is isomorphic to the circle group $U(1)$, so we let the generator $K_3 = (-i/2)\sigma_3$ span the $\mathfrak{u}(1)$ subalgebra -- the reason for this identification is to simplify the later derivation of the gauge potential.

In the case of the flat plane $\Sigma_0^2=\mathbb{R}^2$ (for $k=0$), the quotient works slightly differently: since $\mathbb{R}^2$ is a normal subgroup of $E(2)$ but $O(2)$ is not, we find that $O(2)\simeq E(2)/\mathbb{R}^2$ (through the usual map that associates to an affine transformation a linear transformation of the tangent vector space associated to the affine space). Again though, $U(1)\leq O(2)\leq E(2)$, and we let the generator $K_3$ (in this case, the rotation generator) span the $\mathfrak{u}(1)$ subalgebra. With this identification the remaining two generators (the translations) span the $\mathbb{R}^2$ plane. Using the geometrical construction given above, it can be seen how this continues over from the spherical case, with the $z$-axis rotation generator being $K_3$ in both cases. One can even consider the above spherical case, and simply let the radius of the sphere go to infinity; in that case, the rotation about the $x$- and $y$-axes become the $y$ and $x$ translation generators respectively and the rotation generator stays as it is.

For the hyperbolic space $H^2\equiv\Sigma_{-1}^2$, and $H^2\simeq SU(1,1)/U(1)$. Thus, although it is harder to visualise, we have essentially a similar situation as for $k=1$: one of the generators ($K_3$) is of an ordinary rotation type, and it is this generator that we use to span the subalgebra $\mathfrak{u}(1)$ in the quotient. The other two generators will be hyperbolic rotations on the space $\Sigma_{-1}^2$.

\subsection{The general case}

It can be seen from the previous subsection that we could express this information in a much simpler way, exploiting the three values we have chosen for $k$. Examining (\ref{k1met}, \ref{k0met}, \ref{k-met}), the metric in all the above cases can be expressed as
\begin{equation}\label{genmet}
d\Omega_k^2 = d\theta^2+f_k^2(\theta)d\phi^2
\end{equation}
where
\begin{equation}\label{fkdef}
f_k(\theta) = \left\{ \begin{array}{ll}
\sin\theta & \mbox{for } k=1 \\
\theta & \mbox{for } k=0 \\
\sinh\theta & \mbox{for } k=-1 \end{array}\right.
\end{equation}
Examining (\ref{k1cr}, \ref{k0cr}, \ref{k-cr}), we can also express the commutation relations of the three Lie algebras in the following succinct way:
\begin{equation}\label{gencr}
\begin{array}{ccccc}
[K_1,K_2]=kK_3 & \quad &
[K_2,K_3]=K_1 & \quad &
[K_3,K_1]=K_2
\end{array}
\end{equation}
with the algebra (which we call $\mathfrak{lg}(2)$ in general) being $\mathfrak{su}(2)$ for $k=1$, $\mathfrak{e}(2)$ for $k=0$ and $\mathfrak{su}(1,1)$ for $k=-1$; and where the three generators of the group are the three $K_j$s.

It can be noted that in each of the three cases above, $K_3$ in particular takes a similar form: it has an identical representation each time in the form of differential operators, and thus represents in each case a rotation which can easily be arranged about the same axis (essentially $z$). The significance of this in later sections is linked to the presence of the factor of $k$ which accompanies the $K_3$ generator above. 

Finally, we can also give the topology of the foliated hypersurfaces $\Sigma_k^2$ as 
\begin{equation}
\Sigma_k^2\simeq LG(2)/U(1)
\end{equation}
for $k = \pm 1$ and
\begin{equation}
O(2)\simeq E(2)/\Sigma_k^2
\end{equation}
for $k=0$; and the topology of the 4D manifolds themselves as
\begin{equation}\label{top4man}
\mathbb{R}^2\times\Sigma_k^2;
\end{equation}
where $LG(2)$ is the Lie group in question, and where it is understood that the generator $K_3$ (as noted above) plays a privileged role in forming this quotient.

These facts, and equations (\ref{genmet}, \ref{gencr}) will be useful when later we derive the gauge potential using a proof analogous to that of K\"{u}nzle \cite{21}, and when we derive the field equations themselves.

\section{Topologically non-trivial $\mathfrak{su}(N)$ Einstein-Yang-Mills theory}

In this section we give the general mathematical background we shall need to describe and model topological black holes. The discussion proper will begin with a detailed outline of our metric and gauge potential ans\"{a}tze. This will include a full derivation of the gauge potential in this case, since this is a fairly technical procedure. These ans\"{a}tze will then be used to derive the field equations and boundary conditions for the topological black hole solutions. Finally, we describe some trivial embedded solutions of the theory (in analogy with \cite{1}).\\

\subsection{Ans\"{a}tze}

The action used for the four-dimensional $\mathfrak{su}(N)$ EYM theory with a negative cosmological constant is
\begin{equation}
S_{EYM}=\frac{1}{2}\int d^{4}x\sqrt{-g}[R-2\Lambda-\mbox{Tr}F_{\mu\nu}F^{\mu\nu}],
\end{equation}
where $R$ is the Ricci scalar of the geometry and $\Lambda$ is the cosmological constant. Throughout the paper the metric has signature $(-, +, +, +)$ and we use units in which $4\pi G=1=c$. In this paper, we focus on $\Lambda<0$. Varying the action gives the field equations
\begin{equation}\label{origFE}
\begin{split}
-2T_{\mu\nu}&=R_{\mu\nu}-\frac{1}{2}g_{\mu\nu}R + \Lambda g_{\mu\nu},\\
0&=\nabla_\lambda F^\lambda_{\:\:\mu}+[A_\lambda,F^\lambda_{\:\:\mu}]\\
\end{split}
\end{equation}
where the YM stress-energy tensor is
\begin{equation}\label{origYMseten}
T_{\mu\nu}=\mbox{Tr}F_{\mu\lambda}F_\nu^\lambda-\frac{1}{4}g_{\mu\nu}\mbox{Tr}F_{\lambda\sigma}F^{\lambda\sigma},
\end{equation}
and where it should be noted that `Tr' represents the trace in the Lie algebra sense, rather than the ordinary matrix trace. In equations (\ref{origFE}, \ref{origYMseten}) we have employed the usual Einstein summation convention where it is understood that summation occurs over repeated indices. However where appropriate, summations will be shown explicitly.

In this paper we are interested in static, topological black hole solutions of the field equations \eqref{origFE}, specifically for spaces regularly foliated by 2D (spacelike) hypersurfaces of constant and unit- or zero-magnitude Gaussian curvature $k$, and hence (following the last section) we write the metric in standard Schwarzschild co-ordinates as
\begin{equation}
ds^2=-\mu S^2dt^2+\mu^{-1}dr^2+r^2d\theta^2+r^2f^2_k(\theta)d\phi^2,
\end{equation}
where $\mu$ and $S$ depend on $r$ alone. For convenience, we may take 
\begin{equation}
\mu(r)=k-\frac{2m(r)}{r}-\frac{\Lambda r^2}{3}.
\end{equation}
Note the presence of our modified Gaussian curvature constant $k$. We emphasise again that in our case, $\Lambda<0$; and that the $k=1$ case has been thoroughly investigated in our previous work \cite{1}, and so will be passed over more quickly and mainly used for comparison to the other cases.

The most general gauge potential we used in the spherically symmetric case was \cite{21}:
\begin{equation}\label{S2pot}
A=\mathcal{A}\,dt+\mathcal{B}\,dr+\frac{1}{2}(C-C^H)d\theta-\frac{i}{2}[(C+C^H)\sin\theta+D\cos\theta]d\phi,
\end{equation}
where $\mathcal{A},\mathcal{B},C$ and $D$ are all ($N\times N$) matrices and $C^H$ is the Hermitian conjugate of $C$; the matrices $\mathcal{A}$ and $\mathcal{B}$ are purely imaginary, diagonal and traceless and depend only on $r$; the matrix $C$ (which also depends solely on $r$) is upper-triangular, with non-zero entries only immediately above the diagonal, i.e.:
\begin{equation}\label{Cdef}
C_{j,j+1}=\omega_j(r)e^{i\gamma_j(r)}
\end{equation}
for $j=1,...,N-1$; and $D$ is the constant matrix
\begin{equation}\label{Ddef}
D=\mbox{Diag}(N-1, N-3,...,-N+3, -N+1).
\end{equation}
However in the cases $k=0, -1$ we must alter \eqref{S2pot} slightly to take into account the different geometry. The new gauge potential becomes
\begin{equation}\label{GenGP}
A=\mathcal{A}\,dt+\mathcal{B}\,dr+\frac{1}{2}(C-C^H)d\theta-\frac{i}{2}\left[(C+C^H)f_k(\theta)+D\frac{df_k(\theta)}{d\theta}\right]d\phi,
\end{equation}
with the same definitions for the matrices as above, and including $f_k(\theta)$ (from \eqref{fkdef}) and its first derivative. We shall now demonstrate that this is an appropriate choice of potential, given the symmetry considerations, by using an argument analogous to one due to K\"{u}nzle in \cite{21}.

\subsection{Deriving the ansatz for the gauge potential}

The gauge potential for any system is never unique (see e.g. \cite{21}). By definition it is invariant under the action of some group, so it always has freedom under transformations within that group (automorphisms); therefore it is merely a case of finding one invariant under actions of the requisite symmetry group that fulfills the appropriate gauge constraints -- we shall show that \eqref{GenGP} is such a potential. The procedure to find \textit{all} such irreducible representations can be found in \cite{56}.

Following on from section \ref{2dcc}, we now present the derivation of the gauge potential in the case of general $k$. The similarity in the derivations for the three values of $k$ is such that we can talk very generally about the Lie groups/algebras we are using, as long as we are using one of the three we have specified; so we shall talk about the Lie algebra $\mathfrak{lg}(2)$, realising that this will either mean $\mathfrak{su}(2)$, $\mathfrak{e}(2)$ or $\mathfrak{su}(1,1)$ as appropriate.

What we wish to do is find the possible $LG(2)$-invariant connections on an $SU(N)$ principal bundle $P$ over our spacetime manifold $M$; i.e. find a gauge potential that is invariant under an action of $LG(2)$ by principal bundle automorphisms. According to Wang's theorem \cite{31} (see also \cite{32}), the $LG(2)$-invariant connections on $P$ are in one-to-one correspondence with the linear maps $\mathfrak{lg}(2)\rightarrow \mathfrak{su}(N)$ satisfying $\Pi(X) = \lambda^\prime(X)$ for $X\in \mathfrak{u}(1)$, $\lambda$ being the homomorphism
\begin{equation}\label{u1tosun}
\lambda: U(1)\rightarrow SU(N)
\end{equation}
and $\lambda^\prime$ the induced map of the Lie algebras. Also, we have
\begin{equation}\label{wang1}
\Pi \circ ad_z = ad_{\lambda(z)} \circ\Pi\quad\forall z\in U(1).
\end{equation}
Now, because $SU(2)$ and $SU(1,1)$ have nice succinct matrix representations, but $E(2)$ does not, that means that we have to treat the cases for $k=\pm 1$ very slightly differently to the case $k=0$, but the differences are largely technical. In practice, all it really means is that we keep track of which representation we are using and exactly what is meant by `commutator brackets' when they appear -- whether they represent the obvious matrix commutator, or are the more abstract commutator relationships possessed by any Lie algebra.

Following \cite{21}, let $K_l\quad(l\in\{1,2,3\})$, defined in terms of the general generators referred to in \eqref{gencr}, be a basis for $\mathfrak{lg}(2)$ (i.e., for $k=1$ let $K_l = -(i/2)\sigma_l$; for $k=-1$ let them be the matrices in \eqref{su11map}; and for $k=0$ let them be the more abstract differential operators \eqref{kdiff}). In particular, let the rotation generator $K_3$ in each case span the $\mathfrak{u}(1)$ subalgebra. Note that this associates $K_3$ with the matrix
\begin{equation}
-\frac{i}{2}\left(\begin{array}{cc}
1 & 0\\
0 & -1 \end{array}\right)
\end{equation}
for the cases of $k=\pm 1$, and for the case of $k=0$, with the differential operator stated in \eqref{kdiff}
\begin{equation}
y\frac{\partial}{\partial x} - x\frac{\partial}{\partial y}.
\end{equation}
Now we need a map embedding $U(1)$ isomorphically in $LG(2)$. As was stated in \cite{21}, the map
\begin{equation}\label{zmap}
z\mapsto \left( \begin{array}{cc}
z & 0 \\
0 & z^{-1}\\ 
\end{array} \right)
\end{equation}
embeds $U(1)$ isomorphically in $SU(2)$. Hence, the homomorphisms described by \eqref{u1tosun} are explicitly given as $\lambda(z)=\mbox{diag}(z^{q_1},...,z^{q_N})$ (for sets of $N$ integers $q_i$ such that $\sum^N_i{q_i}=0$, conjugate homomorphisms being described by different orderings of the $q_i$s); and the induced map of Lie algebras is given as $\lambda^\prime_q=-(i/2)D_{q_i}$ where
\begin{equation}\label{dqi}
D_{q_i}=\mbox{diag}(q_1,q_2,...,q_N).
\end{equation}
Obviously we now need to produce analogous maps for $k\neq 1$.

The reason it works in the $k=1$ case is that we can express a general element of $SU(2)$ as the matrix 
\begin{equation}
A=\left( \begin{array}{cc}
a & b \\
-b^* & a^*\\ 
\end{array} \right),
\end{equation}
for some choice of $a,b\in\mathbb{C}$ for which $\mbox{det A} = |a|^2+|b|^2=1$; and then if we let $a=e^{i\theta}=z$ and $b=0$, we find that the matrix specified in \eqref{zmap} is indeed in $SU(2)$, and it reduces infinitesimally to the rotation generator $K_3$. It is for a similar reason that the same map \eqref{zmap} can also be used in the case $k=-1$, since a general element of $SU(1,1)$ can be expressed as 
\begin{equation}
B=\left( \begin{array}{cc}
a & b \\
b^* & a^*\\ 
\end{array} \right),
\end{equation}
for some choice of $a,b\in\mathbb{C}$ for which $\mbox{det B}=|a|^2-|b|^2=1$; and then if we again let $a=e^{i\theta}=z$ and $b=0$, we find that the matrix specified in \eqref{zmap} is also in $SU(1,1)$, and again it reduces infinitesimally to $K_3$.

For $k=0$, it is slightly different, since we are working with $K_3$ as a differential operator. However we are still identifying $K_3$ with the rotation generator of the space. So, given that $z\in U(1)$, we instead simply use the map
\begin{equation}
z\mapsto R_\theta,
\end{equation} 
where $R_\theta$ refers to the rotation generator of $E(2)$ (expressed explicitly in \eqref{ktrans}). This is a map that embeds $U(1)$ isomorphically in $E(2)$, so that once again $\lambda(z)=\mbox{diag}(z^{q_1},...,z^{q_N})$ and $\lambda^\prime_q=-(i/2)D_{q_i}$, as in \eqref{dqi}. We note that infinitesimally, $R_\theta$  can be expressed as the differential operator \eqref{kdiff}, which we associate again with the generator $K_3$, since it obeys the correct commutation laws \eqref{k0cr}.

Defining $\Pi_q\equiv\Pi(K_q)$, then $\Pi_3=-(i/2)D_{q_i}$ and infinitesimally (i.e. where $ad_z:z\mapsto[z,\quad]$) \eqref{wang1} becomes
\begin{equation}\label{wang2}
\Pi([K_3,K_l])=-(i/2)[D_{q_i},\Pi_l]
\end{equation}
where in the second step we have let $z\mapsto K_3$ since this generator is spanning the $\uone$ subalgebra, and hence we replace the map $\lambda$ with the induced map $\lambda'$; and where the square bracketed expression on the right hand sides (but not the left) are the ordinary matrix commutators\footnote{Once again we must note a minor technicality, due to the differing values of $k$ which necessitates a slightly different treatment of the map embedding $U(1)$ in $SU(2)$. For $k=\pm 1$, the square brackets on the left of \eqref{wang2}, i.e. the abstract commutator relations, become the ordinary matrix commutator in the case that the representation used is the matrix representation given in terms of either the Pauli matrices (for $k=1$) or the matrices shown in \eqref{su11map} (for $k=-1$). For $k=0$ however, the square brackets refer to the more abstract commutator, as defined in \eqref{k0cr}, acting on the differential expressions \eqref{kdiff}.}.) Using the commutation relations \eqref{gencr}, we find that
\begin{equation}
\begin{array}{ccc}
\Pi_1=(i/2)[D_{q_i},\Pi_2], & \quad & 
\Pi_2=-(i/2)[D_{q_i},\Pi_1].
\end{array}
\end{equation}
Substituting the second of those expressions into the first gives us
\begin{equation}
\Pi_1=(1/4)[D_{q_i},[D_{q_i},\Pi_1]],
\end{equation}
from which it is easy to see that $((q_i-q_j)^2-4)\Pi_{1,ij}=0$. Given this, and since the $\Pi$ are traceless and anti-Hermitian (and assuming that $q_i \geq q_j$ for $i < j$), we can deduce that
\begin{equation}\label{lambdaCforms}
\begin{array}{cc}
\Pi_1=\frac{1}{2}(C-C^H), & \Pi_2=-\frac{i}{2}(C+C^H),\\
\end{array}
\end{equation}
where $C$ is (as previously described) an upper triangular complex $N\times N$ matrix with Hermitian conjugate $C^H$ and where $C_{ij}\neq 0$ iff 
\begin{equation}\label{kijconstr}
q_i = q_j+2.
\end{equation}
According to the second part of Wang's theorem, the curvature 2-form of this invariant connection on the two-dimensional subspace is given by
\begin{equation}
\langle\tilde{X}\wedge\tilde{Y}/\Omega\rangle=[\Pi(X),\Pi(Y)] - \Pi([X,Y])
\end{equation}
where $X, Y\in SU(N)$, and $\tilde{X}$ and $\tilde{Y}$ are the corresponding generators of $P$ -- since $\Omega=d\tilde{A}+\tilde{A}\wedge\tilde{A}$ here refers purely to the angular part of the gauge potential 1-form $\tilde{A}$, then $\langle\tilde{X}\wedge\tilde{Y}/\Omega\rangle$ represents the curvature 2-form on the bundle itself restricted to the angular part. Therefore, we find
\begin{equation}
\langle\tilde{K_1}\wedge\tilde{K_2}/\Omega\rangle=[\Pi_1,\Pi_2] - \Pi\left([K_1,K_2]\right).
\end{equation}

Substituting in the forms for $\Pi_1$ and $\Pi_2$ given in \eqref{lambdaCforms}, and noting from \eqref{gencr} that (since $\Pi$ is a linear map)
\begin{equation}
\Pi([K_1,K_2]) = -(i/2)kD_{q_i}
\end{equation}
we find that the curvature is given as
\begin{equation}
\tilde{F}\equiv\langle\tilde{K_1}\wedge\tilde{K_2}/\Omega\rangle=-(i/2)([C,C^H]-kD)
\end{equation}
with other components vanishing.

However, since the curvature is a tensorial form it must be the pullback of a scalar multiple of the area element of the 2-space we are working with (sphere, plane or hyperboloid), which we have already described metrically using the function $f_k(\theta)$. That is, the angular component of the full 4D curvature must be of the form $\tilde{F}f_k(\theta)d\theta\wedge d\phi$ (in the $(\theta,\phi)$ co-ordinate system appropriate to the space in question.) This gives us the equation
\begin{equation}\label{struc}
d\tilde{A}+\frac{1}{2}[\tilde{A}, \tilde{A}]=\tilde{F}f_k(\theta)d\theta\wedge d\phi.
\end{equation}
Now we are ready to construct the angular part of the connection, i.e. the local potential $\tilde{A}=A_\theta d\theta + A_\phi d\phi$, an $\mathfrak{su}(N)$ valued one-form on the two dimensional subspace. We can easily check that \eqref{struc} is satisfied by the potential
\begin{equation}
\begin{array}{rlcrl}
A_\theta&=\frac{1}{2}(C-C^H), & \quad & 
A_\phi&=-\frac{i}{2}\left[ (C+C^H)f_k(\theta)+D\frac{df_k}{d\theta}\right].
\end{array}
\end{equation}
A helpful relation in showing this is
\begin{equation}
\frac{d^2f_k(\theta)}{d\theta^2}=-kf_k(\theta)\quad\forall k\in \{-1,0,1\};
\end{equation}
note that the presence of the $\frac{df_k}{d\theta}$ term is due to the fact that the curvature must be the pullback of a tensorial form and hence we end up with a logarithmic derivative, evident when we take the $f_k(\theta)$ out of the square brackets in $A_\phi$. 

Finally, we augment the potential with $t$ and $r$ components, which we may do easily due to the product topology of the whole 4-manifold \eqref{top4man}, such that
\begin{equation}
\begin{array}{l}
A_t\equiv\mathcal{A}\,dt,\\
A_r\equiv\mathcal{B}\,dr,
\end{array}
\end{equation}
where the forms of $\mathcal{A}$ and $\mathcal{B}$ are given in the next subsection.

Hence we have confirmed our ansatz for the gauge potential \eqref{GenGP}. As we noted, the gauge potential is never unique since there are degrees of freedom left over in the structure equations -- there are many other possible forms, including Witten's \cite{33} -- but it is enough that we have found a manifestly $SU(N)$ invariant gauge potential that satisfies the structural constraint \eqref{struc}. Another approach that can be taken is to use the symmetry equations for the Lie algebra in question, as in \cite{56}.

\subsection{Deriving the field equations}

We now outline a way of constructing an irreducible representation of $\mathfrak{su}(N)$ \cite{21} with reference to the previously derived gauge potential \eqref{GenGP}.

We first consider $\mathcal{A}$ and $\mathcal{B}$ to be general imaginary diagonal and traceless
matrices which can be written
in the form
\begin{equation}
\begin{array}{cc}
\left(\mathcal{A}\right)_{jj}=\frac{i}{2}\alpha_{j}(r),\quad & \left(\mathcal{B}\right)_{jj}=\frac{i}{2}\beta_{j}(r)\\
\end{array}
\end{equation}
for real functions $\alpha_j(r)$, $\beta_j(r)$.

We also recall that the form of the diagonal matrix $D$ is specified uniquely, given the equation \eqref{Ddef}, i.e.
\begin{equation}
\left(D\right)_{jj}=N+1-2j.
\end{equation}

The form of the matrices $C$ (and hence $C^H$) were also determined in the previous section (\ref{Cdef}, \ref{Ddef}). Now, we can simplify the field equations considerably:

\begin{enumerate}
\item For purely magnetic solutions, we must set $\alpha_j=0$ (as opposed to the cases where dyons are considered -- see for instance \cite{13, 14, 50}). 
\item Due to gauge freedom, we may set $\beta_j=0$ as well.
\item Finally, one of the Yang-Mills equations has the immediate solution $\gamma_j=0$.
\end{enumerate}
The last item also gives us a final form for $C$ -- the only non-zero elements, indexed by $(j, j+1)$, are

\begin{equation}
(C)_{j,j+1}=\omega_j.
\end{equation}
The details of this derivation are given in \cite{21, 25}.\par The Yang-Mills equations with a cosmological constant thus take the form:
\begin{equation}\label{FEsunkom}
r^2\mu\omega''_{j}+\left(2m-2r^3 p_{\theta,k}-\frac{2\Lambda
r^3}{3}\right)\omega'_{j}+W_{k,j}\omega_j=0
\end{equation}
where
\begin{equation}
\begin{split}
p_{\theta,k}&=\frac{1}{4r^4}\sum^N_{j=1}\left[\left(\omega^2_j-\omega^2_{j-1}-k(N+1-2j)\right)^2\right],\\
W_{k,j}&=k-\omega^2_j+\frac{1}{2}\left(\omega^2_{j-1}+\omega^2_{j+1}\right),
\end{split}
\end{equation}
and the Einstein equations take the form
\begin{equation}\label{FEsunkmpr}
m' =\mu G+r^2p_{\theta,k}
\end{equation}
and
\begin{equation}\label{FEsunkd}
\Delta' = \frac{S'}{S}=\frac{2G}{r},
\end{equation}
where
\begin{equation}
G=\sum^{N-1}_{j=1}\omega_j'^2.\\
\end{equation}
It can readily be checked that these reduce in the correct limit -- see equations (6, 7, 8) in \cite{2}. We can note several things.

\begin{enumerate}
\item From now on, we assume that all the $\omega_j(r)$ are in general non-zero so that we are dealing with genuinely $\mathfrak{su}(N)$ solutions (see, for example, \cite{27, 28, 29, 30} for the consequences of violating this assumption in the $\Lambda=0$ case).

\item For later convenience, we have also introduced $S\equiv e^\Delta$. 

\item We define $\omega_0\equiv\omega_N\equiv 0$, so that (\ref{FEsunkom}, \ref{FEsunkmpr}, \ref{FEsunkd}) reduce to the $\mathfrak{su}(2)$ equations in the correct limit (see below).

\item As in the purely $k=1$ case, there are two symmetries respected by the ans\"{a}tze. Firstly the mapping $\omega_j\rightarrow -\omega_j$ is an invariant mapping separately for each $j$, and secondly the substitution $j\rightarrow N-j$ for all $j$ is a symmetry. 

\end{enumerate}

Finally we note that, since the Einstein equation for $S$ decouples from the rest, it can be integrated separately once we know the character of the metric function $\mu$ and the gauge field functions $\omega_j$, to give the solution
\begin{equation}
S=e^{\Delta}=\exp\left(\int^r_c\frac{2G}{r}dr\right)
\end{equation}
for some arbitrary constant $c$. So the $N$ functions on which we will
concentrate are $\mu$ and the $\omega_j$s. 

For completeness, we shall show what forms these equations take for the simplest non-trivial case, $\mathfrak{su}(2)$. In this case, there is one gauge field function which we call $\omega$, and the following functions simplify considerably:
\begin{equation}\label{FEsu2komdefs}
\begin{split}
p_{\theta,k}&=\frac{(k-\omega^2)^2}{4r^4},\\
G&=\omega'^{2}.
\end{split}
\end{equation}
This means the field equations \eqref{FEsunkom} become the single equation
\begin{equation}\label{FEsu2kom}
r^2\mu\omega''+\left(2m-2r^3p_{\theta,k}-\frac{\Lambda r^3}{3}\right)\omega'+(k-\omega^2)\omega=0,
\end{equation}
and the field equations (\ref{FEsunkmpr}, \ref{FEsunkd}) become
\begin{equation}\label{FEsu2kmprd}
\begin{split}
m' &=\mu\omega'^2+\frac{(k-\omega^2)^2}{4r^2},\\
\Delta' &= \frac{S'}{S}=\frac{2G}{r}=\frac{2\omega'^2}{r}.
\end{split}
\end{equation}
The existence of exact solutions to these $\mathfrak{su}(2)$ equations will be necessary later in our constructive proof of the existence of non-trivial solutions to the $\mathfrak{su}(N)$ equations.

\subsection{Boundary conditions}\label{secbcs}

We are interested in black hole solutions to the field equations (\ref{FEsunkom}, \ref{FEsunkmpr}, \ref{FEsunkd}), but these equations are singular at the event horizon $r=r_h$ and as $r\rightarrow\infty$. (Since $\Lambda<0$, there is no cosmological horizon to consider.) A first step is to derive the boundary conditions at these singular points. Local existence has been proven already in $\mathfrak{su}(N)$ EYM theory for the cases $\Lambda=0$ \cite{24, 25} and for $\Lambda<0$ with spherical symmetry ($k=1$) \cite{1}; our intention is to extend these  results to $k\neq 1$.

We are ignoring solitons, the particle-like solutions to the equations which possess no event horizon and so continue to be regular in the limit $r\rightarrow0$. (See \cite{1, 4, 5, 9, 11, 38, 39} among others, for a discussion of these in the literature.) The Ricci curvature scalar $R = g_{\mu\nu}R^{\mu\nu}$ (for $R^{\mu\nu}$ the Ricci tensor) is complicated, but always involves the terms:
\begin{equation}\label{riemannscalar}
R=-4\Lambda+\frac{2(k-1)}{r^2}+...
\end{equation}
where the ellipses refer to terms involving $m$, $S$ and their derivatives. Hence, only for $k=1$ is the curvature non-singular at $r=0$. For $k\neq 1$, at $r=0$ the Riemann scalar has an essential singularity, hence there is no such thing as globally regular solutions in these cases. Hence, we shall only need to pay attention to (exterior) black hole solutions, meaning that our two singular points of interest are the event horizon $r=r_h$ and the asymptotic region $r\rightarrow\infty$.

\subsubsection{Event horizon}

We assume that the black hole solutions have a regular, non-extremal event horizon at $r=r_h$; i.e. for which $\mu(r_h)$ has a single zero. This fixes the value of $m_h\equiv m(r_h)$ as:
\begin{equation}\label{2mrhk}
2m_h=kr_h-\frac{\Lambda r_h^3}{3}.
\end{equation}

We assume that the field variables $m(r)$, $S(r)$ and $\omega_j(r)$ have regular Taylor expansions about $r_h$, i.e. that for $r\approx r_h$,
\begin{equation}\label{expanrh}
\begin{split}
m(r)&=m(r_h)+m'(r_h)(r-r_h)+O(r-r_h)^2,\\
S(r)&=S(r_h)+S'(r_h)(r-r_h)+O(r-r_h)^2,\\
\omega_j(r)&=\omega_j(r_j)+\omega'_j(r_h)(r-r_h)+O(r-r_h)^2.\\
\end{split}
\end{equation}
Letting $\mu(r_h)=0$ in the field equation \eqref{FEsunkom} gives us the following boundary conditions for $\omega'_j(r_h)$:
\begin{equation}\label{om'rhk}
\omega'_j(r_h)=\frac{W_{k,j}(r_h)\omega_j(r_h)}{2m(r_h)-2r^3_hp_{\theta,k}(r_h)-\frac{2\Lambda r_h^3}{3}}.
\end{equation}
So for fixed $\Lambda$ the expansions \eqref{expanrh} are determined entirely by the $N+1$ quantities given by $r_h$, $\omega_j(r_h)\equiv\omega_{j,h}$ and $S(r_h)$.
Setting $\mu(r_h)=0$ in the field equation \eqref{FEsunkmpr}, we can show that for a non-extremal event horizon we require
\begin{equation}\label{2m'rhk}
2m'(r_h)=2r^2_hp_{\theta,k}(r_h)<k-\Lambda r_h^2,
\end{equation}
which weakly constrains the possible values of $\omega_j$ near the horizon. It also follows from \eqref{2m'rhk} that for $k=-1$, we have a minimum event horizon radius given by the constraint
\begin{equation}\label{minrhk-1}
r_h>\sqrt{\frac{-1}{2p_{\theta,k}(r_h)+\Lambda}}.
\end{equation}
Since the field equations are invariant under $\omega_j\mapsto -\omega_j$, we may consider $\omega_{j,h}>0$ (without loss of generality).

\subsubsection{Infinity}

At infinity, we expect that the black hole will approach the topological analogue to adS space, i.e. that
\begin{equation}
\mu(r)\rightarrow k-\frac{\Lambda r^2}{3}
\end{equation}
as $r\rightarrow\infty$. We assume that all the field variables have regular expansions in $r^{-1}$ as $r$ approaches infinity:
\begin{equation}\label{expaninf}
\begin{array}{rclcrcl}
m(r)&=&M+O(r^{-1}),&\qquad&S(r)&=&1+O(r^{-1}),\\
\omega_j(r)&=&\omega_{j,\infty}+O(r^{-1}).&&&&\\
\end{array}
\end{equation}
Just as in the spherically symmetric case with $\Lambda<0$, there are no \textit{a priori} constraints on the values of $\omega_{j,\infty}$, so that in general, the adS topological black holes will carry a global magnetic charge \cite{49}.

\subsection{Trivial solutions}

A key part of our proof of the existence of non-trivial $\mathfrak{su}(N)$ solutions is the analyticity in the initial conditions of the field variables, so that we may find solutions with initial conditions in a neighbourhood of those producing known solutions. Thus, it is important that we can find some `trivial' known solutions.

The field equations (\ref{FEsunkom}, \ref{FEsunkmpr}, \ref{FEsunkd}) are non-linear coupled equations, but they possess a number of analytic, trivial solutions. These arise by letting the functions $\omega_j(r)$ be identical to a constant. This produces as a constraint 
\begin{equation}\label{wcon}
W_{k,j}\omega_j=0.
\end{equation}

\subsubsection{Reissner-Nordstr\"{o}m-adS (RNadS)}

Equation \eqref{wcon} clearly has the solution $\omega_j(r)\equiv 0$ for all $j$, which gives the RNadS black hole with metric function
\begin{equation}
\mu(r)=k-\frac{2M}{r}+\frac{Q^2}{r^2}-\frac{\Lambda r^2}{3}
\end{equation}
where the magnetic charge $Q$ is defined as
\begin{equation}
Q^2=\frac{k^2}{6}N(N+1)(N-1).
\end{equation}
This solution exists for all three values of $k$. Note that the magnetically charged RNadS black hole is only a solution of the field equations if the charge is exactly this value. 

Note also that for $k=0$ there is a difference. In that case, $Q=0$, and hence in the $\mathfrak{su}(N)$ case with planar symmetry, the solution is more similar to the Schwarzchild-adS solution, discussed next.

\subsubsection{Schwarzchild-adS (SadS)}

Writing \eqref{wcon} out in full and letting $\Omega_j\equiv\omega_j^2$ turns the $W_{k,j}=0$ into a system of linear equations, with the solution
\begin{equation}\label{Omsol}
\Omega_j=kj(N-j).
\end{equation}
For $k=1$, we have $\Omega_j=\omega^2_j=j(N-j)$, so that
\begin{equation}
\omega_j=\pm\sqrt{j(N-j)},
\end{equation}
for all $j$, which agrees with previous work \cite{1}. This gives us
\begin{equation}
m(r)=M=\mbox{ constant.}
\end{equation}

For the case $k=-1$, it can be immediately seen that for \eqref{Omsol} to be correct, $\omega_j$ will have to be imaginary. Thus in this case, all we have is RNadS-type solutions (due to $Q\neq 0$) and the only trivial embedded solutions are found by setting $\omega_j\equiv 0$ for all $j$. For $k=0$ however, it is obvious that $\omega_j\equiv 0$, which coincides with the $Q=0$ case above, hence the only type of solution we have is the SadS-type solution.

\subsubsection{$\mathfrak{su}(2)$ embedded solutions}\label{secembed}

We now demonstrate that it is possible to embed any solution of the $\mathfrak{su}(2)$ field equations as an $\mathfrak{su}(N)$ solution by a simple rescaling of variables. Note that this is a fairly trivial step from the similar $\Lambda=0$, $k=1$ case discussed in \cite{25, 34} and the $\Lambda<0$, $k=1$ case discussed in \cite{1}.

\begin{prop}
Any $\mathfrak{su}(2)$ solution can be rescaled and embedded as an
$\mathfrak{su}(N)$ solution.
\end{prop}

\textbf{Proof} We begin with the field equations (\ref{FEsunkom}, \ref{FEsunkmpr}, \ref{FEsunkd}). We attempt to rescale them with the following definitions:
\begin{equation}\label{embedscale}
\begin{array}{lll}
\tilde{N}\equiv\frac{1}{6}N(N-1)(N+1), & R\equiv\tilde{N}^{-\frac{1}{2}}r, & \tilde{\Lambda}\equiv\tilde{N}\Lambda, \\
\omega_j\mapsto\sqrt{j(N-j)}\omega, & \tilde{m}\equiv\tilde{N}^{-\frac{1}{2}}m. & \\
\end{array}
\end{equation}
This rescaling leads to the following equations:
\begin{equation}
\begin{array}{rl}\label{su2embed}
0&=R^2
\mu\left(\disfrac{d^2\omega}{dR^2}\right)+\left(2\tilde{m}-\disfrac{2\tilde{\Lambda}R^3}{3}-2R^3\tilde{p}_{\theta,k}\right)
\left(\disfrac{d\omega}{dR}\right)+(k-\omega^2)\omega\\
&\\
\disfrac{d\tilde{m}}{dR}&=\mu\left(\disfrac{d\omega}{dR}\right)^2+R^2\tilde{p}_{\theta,k}\\
&\\
\disfrac{1}{S}\frac{dS}{dR}&=\disfrac{2}{R}\left(\disfrac{d\omega}{dR}\right)^2\\
\end{array}
\end{equation}
with
\begin{equation}
\tilde{p}_{\theta,k}\equiv\disfrac{(\omega^2-k)^2}{2R^4}.
\end{equation}

It can be checked that these are exactly the same equations as the $\mathfrak{su}$(2) field equations with general $k$ (\ref{FEsu2kom}, \ref{FEsu2kmprd}).$\Box$\\

It is easy to see that the boundary conditions (\ref{expanrh}, \ref{expaninf}) reduce to the topological $\mathfrak{su}(2)$ boundary equations given in \cite{2}. We have therefore proven that any $\mathfrak{su}(2)$ topological black hole solution can be embedded into $\mathfrak{su}(N)$ EYM to yield an asymptotically topological anti-de-Sitter black hole. One final fact of note is that these embedded solutions will have the same number of nodes as the original $\essu$ solution, since if $\omega(r_0)=0$ for some $r_0$, then so will $\tilde{\omega}_j(r_0)=0$, according to \eqref{embedscale}. This fact will be of importance in the proof of our second result (Theorem \ref{neighproof}).

\section{Existence of non-trivial $\mathfrak{su}(N)$ black hole solutions}

Our goal is to prove that we can find solutions throughout the whole range of the spacetime; in other words, solutions which can be proven to exist at the event horizon, and can be integrated out arbitrarily far into the asymptotic regime. The way we do this is by proving that solutions exist locally at the event horizon and locally as $r\rightarrow\infty$; then we prove that given any solution which remains regular in a specified range, we can continue to integrate that solution outwards, right into the asymptotic regime. We use these results to argue that locally regular solutions can be `patched together' to form global black hole solutions. We note existence has already been proven for $\Lambda=0$ case in $\mathfrak{su}(N)$ EYM \cite{34} and with $\Lambda < 0$ and $k=1$ \cite{1}.

\subsection{Local existence of solutions}

In this section, following e.g. \cite{6}, we shall make use of an established
theorem of differential equations \cite{35} to demonstrate the local existence of solutions at the event horizon and near infinity. We begin by stating the theorem:

\begin{thr}\label{exdetheor}\cite{6}
Consider a system of differential equations for $n+m$ functions $\mathbf{a}=(a_1,a_2,\ldots,a_n)$ and $\mathbf{b}=(b_1,b_2,\ldots,b_m)$ of the form
\begin{equation} \label{theorem}
\begin{split}
x\frac{da_i}{dx}&=x^{p_i}f_i(x,\mathbf{a},\mathbf{b}),\\
x\frac{db_i}{dx}&=-\lambda_i
b_i+x^{q_i}g_i(x,\mathbf{a},\mathbf{b})\\
\end{split}
\end{equation}
with constants $\lambda_i>0$ and integers $p_i,q_i\geq1$ and let $\mathcal{C}$ be an open subset of $\mathbb{R}^n$ such that the functions $f_i$ and $g_i$ are analytic in a neighbourhood of $x=0$, $\mathbf{a}=\mathbf{c}$, $\mathbf{v}=\mathbf{0}$, for all  $\mathbf{c}\in\mathcal{C}$. Then there exists an $n$-parameter family of solutions of the system such that
\begin{equation}\label{bcs}
\begin{matrix}
&a_i(x)=c_i+O(x^{p_i}),&b_i=O(x^{q_i}),
\end{matrix}
\end{equation}
where $a_i(x)$ and $b_i(x)$ are defined for $\mathbf{c}\in\mathcal{C}$, 
$|x|<x_0(\mathbf{c})$ and are analytic in $x$ and $\mathbf{c}$.
\end{thr}

This theorem allows us to parameterize the family of solutions near a singular point of a set of ordinary differential equations. We need to take each singular point in turn (here, $r=r_h$ and $r\rightarrow\infty$) and change variables so that the field equations are in the form required by the theorem. After that, it is elementary to verify the forms we have chosen for our expansions of the field variables near the singular points (\ref{expanrh}, \ref{expaninf}).

\subsubsection{Local existence of solutions at the event horizon}

Now we turn our attention to the field equations at $r=r_h$. We assume the existence of a non-degenerate event horizon, so that $\mu(r_h)=0$ but $\mu'(r_h)>0$ is finite. We begin by proving a proposition analogous to that proven for $\mathfrak{su}(N)$, $k=1$ in \cite{1}.

\begin{prop}\label{lexeh}
\label{exrh}
There exists an N-parameter family of local solutions of the field
equations near $r=r_h$ analytic in $r_h$, $\omega_{j,h}$ and
$\rho=r-r_h$ such that
\begin{equation}
\begin{split}
\mu(r_h+\rho)&=\mu'(r_h)+O(\rho),\\
\omega_j(r_h+\rho)&=\omega_{j,h}+\omega'(r_h)\rho+O(\rho^2),\\
\end{split}
\end{equation}
where $\mu'(r_h)$ and $\omega'(r_h)$ are functions of the
$\omega_{j,h}$.
\end{prop}

\textbf{Proof} Following \cite{6, 34, 24, 1}, let us define a new independent variable $x=r-r_h$, and define
some new dependent variables:
\begin{equation}
\rho\equiv
r,\quad\quad\lambda\equiv\frac{\mu}{\rho},\quad\quad\psi_j\equiv\omega_j,\quad\quad\xi_j\equiv\frac{\mu\omega'_j}{\rho}=\lambda\omega_j'.\\
\end{equation}
The field equations take the form:
\begin{equation}\label{rhlocalex}
\begin{array}{lll}
x\disfrac{d\rho}{dx}=x, & x\disfrac{d\psi_j}{dx}=\disfrac{x\xi_j}{\lambda}, & x\disfrac{d\lambda}{dx}=-\lambda+xH_\lambda+F_\lambda,\\
&&\\
x\disfrac{d\Lambda}{dx}=0, & x\disfrac{dS}{dx}=\disfrac{2x}{\rho}GS, & x\disfrac{d\xi_j}{dx}=-\xi_j+xH_{\xi_j}+F_{\xi_j};
\end{array}
\end{equation}
where
\begin{equation}
\begin{split}
F_\lambda\equiv &\frac{1}{\rho}\left(k-\frac{2}{\rho^2}\mathcal{P}-\Lambda\rho^2\right),\\
H_\lambda\equiv &-\frac{\lambda}{x}(1+2\mathcal{G}),\\
F_{\xi_j}\equiv &-\frac{W_{k,j}\psi_j}{\rho^2},\\
H_{\xi_j}\equiv &x\frac{H_\lambda\xi_j}{\lambda};\\
\end{split}
\end{equation}
and we note that
\begin{equation}
\begin{split}\label{wpsi}
W_{k,j}=&k-\psi^2_j+\frac{1}{2}(\psi^2_{j-1}+\psi^2_{j+1});
\end{split}
\end{equation}
and define
\begin{equation}\label{Ppsi}
\begin{split}
\mathcal{P}\equiv & r^4p_{\theta,k}=\frac{1}{4}\sum^N_{j=1}\left[\left(\psi^2_j-\psi^2_{j-1}-k(N+1-2j)\right)^2\right],\\
\mathcal{G}\equiv & \frac{1}{\lambda^2}\sum^{N-1}_{j=1}\xi_j^2.
\end{split}
\end{equation}
So the functions $F_\lambda$, $F_{\xi_j}$, $H_\lambda$ and $H_{\xi_j}$ are polynomials in $1/\rho$, $1/\lambda$, $\rho$, $\lambda$, and $\Lambda$.

The equations \eqref{rhlocalex} are not yet in the required form, so next we let
\begin{equation}
\begin{array}{cc}
\tilde{\xi_j}\equiv\xi_j-F_{\xi_j}, & \tilde{\lambda}\equiv\lambda-F_\lambda.
\end{array}
\end{equation}
This makes the two non-conforming equations (for $xd\xi_j/dx$ and $xd\lambda/dx$) take the form:
\begin{equation}
\begin{split}
x\frac{d\tilde{\xi_j}}{dx}&\equiv-\tilde{\xi_j}+xG_{\xi_j},\\
x\frac{d\tilde{\lambda}}{dx}&\equiv-\tilde{\lambda}+x G_\lambda,\\
\end{split}
\end{equation}
in which $G_\lambda$ and $G_{\xi_j}$ are polynomials given by
\begin{equation}
G_{\xi_j}\equiv H_{\xi_j}-\frac{d F_{\xi_j}}{dx},\quad\quad
G_{\lambda}\equiv H_{\lambda}-\frac{d F_\lambda}{dx}.
\end{equation}

The polynomials $G_\lambda$ and $G_{\xi_j}$ are lengthy; suffice it to say, it can be checked that both are polynomials in $1/\rho$, $1/\lambda$, $\rho$, $\lambda$, $\tilde{\lambda}$, $\tilde{\xi_j}$ and $\Lambda$.
Hence, due to Theorem \ref{exdetheor}, there then exist solutions to the equations of
the form
\begin{equation}
\begin{array}{lll}
\rho=r_h+O(x), & \psi_j=\omega_{j,h}+O(x), & \tilde{\lambda}=O(x), \\ 
\tilde{\xi_j}=O(x), & S=S(r_h)+O(x); & \end{array}
\end{equation}
with $\rho$, $\tilde{\lambda}$, $\psi_j$ and $\tilde{\xi_j}$ all analytic in $x$, $r_h$, $\omega_j(r_h)$, $\Lambda$ and $S(r_h)$. Transforming back to our original variables gives us the correct behaviour and analyticity.$\Box$ \\

We have thus proven existence of solutions to the field equations for a black hole in some neighbourhood of the event horizon $r=r_h$, satisfying the boundary conditions \eqref{expanrh}. 

\subsubsection{Local existence of solutions at infinity}
\label{exinf}

Local existence at infinity for asymptotically adS spherically symmetric black holes has been proven in \cite{1}, and the proof extends almost directly to the topological case. We note that, as in the adS spherically symmetric case, it is relatively easy to prove existence and we need only go to first order in the field variables, unlike in the asymptotically flat case where higher order terms were needed and the analysis was more involved \cite{24, 25}. We also note that for adS space, the field equations are not actually singular as $r\rightarrow\infty$, but it is convenient to use this method anyway -- the above theorem does not require the boundary points to be singular, but it can account for that if needed.

Again we prove a proposition analogous to one in \cite{12}.

\begin{prop}
\label{lexinf}
There exists an $2N$-parameter family of local solutions of the
field equations near $r=\infty$, analytic in $\Lambda$, $\omega_{j,\infty}$, $M$ and $r^{-1}$ such that
\begin{equation}
\begin{split}
\mu(r)&=k-\frac{2M}{r}-\frac{\Lambda r^2}{3}+O\left(\frac{1}{r^2}\right),\\
\omega_j(r)&=\omega_{j,\infty}-\frac{c_j}{r}+O\left(\frac{1}{r^2}\right).\\
\end{split}
\end{equation}
\end{prop}

\textbf{Proof} We introduce new variables, following \cite{6, 25, 1}:
\begin{equation}
\begin{matrix}
&x=r^{-1},&\psi_j=\omega_j,&\xi_j=r^2\omega'_j,&\lambda=r\left(k-\mu-\frac{\Lambda
r^2}{3}\right)\equiv 2m.&
\end{matrix}
\end{equation}
Then the field equations take the form:
\begin{equation}
\begin{array}{lll}
x\disfrac{d\lambda}{dx}=-xf_\lambda, & x\disfrac{d\psi_j}{dx}=-x\xi_j, & 
x\disfrac{d\xi_j}{dx}=xg_{\xi_{j}},\\
&&\\
x\disfrac{dS}{dx}=x^4f_S, & x\disfrac{d\Lambda}{dx}=0; &
\end{array}
\end{equation}
where
\begin{equation}
\begin{array}{rl}
f_S\equiv& S\sum_{j=1}^{N-1}\xi_j^2,\\
f_\lambda\equiv&
2\mu x^2\sum^{N-1}_{j=1}\xi^2_j+\frac{1}{2}\sum^N_{j=1}\left(\psi_j^2-\psi_{j-1}^2-k(N+1-2j)\right)^2,\\
g_{\xi_j}\equiv & -\frac{W_{j,k}\psi_j}{\mu x^2}+\frac{1}{\mu}\left(\lambda-\frac{2\mathcal{P}}{x^3}\right)\\
\end{array}
\end{equation}
(where we have used the forms from (\ref{wpsi}, \ref{Ppsi})); and since 1/$\mu$ is at least of order $x^2$ as $x\rightarrow 0$, it can be observed that all of these polynomials are non-singular as $x\rightarrow 0$. Therefore we have proven local existence of solutions at infinity, and Theorem \ref{exdetheor} confirms that the functions exhibit the required behaviour near infinity \eqref{expaninf}. We also note that solutions are analytic in $x$, $\Lambda$, $M$, $\omega_{j,\infty}$ and $S_\infty$. Also, by rescaling the time co-ordinate in the metric we can fix $S_\infty=1$ so that the spacetime is asymptotically `topological adS' as we have described it -- this satisfies the boundary conditions \eqref{expaninf}, and the field variables are analytic in $M$, $\omega_{j,\infty}$, $c_j$, $r$ and $\Lambda$.$\Box$

\subsection{Global regularity for $\mu>0$}

Our strategy for proving the existence of genuinely $\mathfrak{su}(N)$ solutions involves showing that given initial conditions near the singular point at the event horizon $r=r_h$, we can regularly integrate the field equations out to infinity.
To be specific, we wish to show that as long as $\mu>0$ (required for the spacetime to be regular), then any solution that is regular up to a certain $r$ value can be extended  for larger values of $r$. To do this we use the field equations to prove the following lemma, which has been proven for the spherically symmetric cases of $\mathfrak{su}(2)$ with $\Lambda=0$ \cite{6}, of $\mathfrak{su}(N)$ with $\Lambda=0$ \cite{34}, and of $\mathfrak{su}(N)$ with $\Lambda<0$ and $k=1$ \cite{1}.

\begin{lem}
\label{globreg}
As long as $\mu>0$ all field variables are regular functions of
$r$.
\end{lem}
To prove this lemma we shall take some interval $I=[r_0,r_1)$ (such that $r_h<r_0<r_1$), assume the solution is regular in this interval and that $\mu(r)>0$ on the interval $\bar{I}=[r_0,r_1]$, and then show that this implies regularity at $r=r_1$ as well. In
other words, as long as $\mu>0$, if we start off with arbitrary initial conditions outside the event horizon (i.e. a `piece' of a regular solution) then we can continue to integrate it regularly outwards into the asymptotic region. This
method is adapted from a similar one used in \cite{6}.

\textbf{Proof} If $\mu(r_1)>0$, then by integrating the equation for $m'$ in 
\eqref{FEsunkmpr}, we obtain
\begin{equation}
kr_1-\frac{\Lambda r_1^3}{3}>2m(r_1)\geq
2\int^{r_1}_{r_0}\sum_{j=1}^{N-1}\mu\omega'^2_j dr.
\end{equation}
Since $\mu$ must have a minimum in $\bar{I}$, we define
\begin{equation}
\mu_{min}\equiv\inf(\mu:r\in\bar{I})>0.
\end{equation}
Therefore:
\begin{equation}
2\mu_{min}\int^{r_1}_{r_0}\sum_{j=1}^{N-1}\omega'^2_j dr\leq
kr_1-\frac{\Lambda r_1^3}{3}
\end{equation}
or
\begin{equation}
\int^{r_1}_{r_0}\sum_{j=1}^{N-1}\omega'^2_j
dr\leq\frac{kr_1-\frac{\Lambda r_1^3}{3}}{2\mu_{min}}.
\end{equation}
The LHS is bounded above by the RHS, which we notice is constrained to be positive -- this is because of the constraint \eqref{2mrhk}, where we remember we require $m_h>0$, and because the mass function is monotonic ($m'(r)>0\:\: \forall r$), which together means that $m(r)>0\:\: \forall r$. We can also see that since the LHS is a sum of squared (i.e. positive) terms in $\omega'_j$, each term must be bounded below by zero. Hence $G$ is bounded and so is $2G/r$ (and thus $\Delta'$), so direct integration gives us $\Delta$ (or $S$) bounded. 

Using the Cauchy-Schwarz inequality and performing an integration gives us
\begin{equation}
\begin{split}
\int^{r_1}_{r_0}\sum_{j=1}^{N-1}\omega'^2_jdr&\geq\frac{1}{r_1-r_0}\sum_{j=1}^{N-1}(\omega_j(r_1)-\omega_j(r_0))^2.
\end{split}
\end{equation}
From here we see that each $\omega_j(r_1)$ is finite.

Finally, we take the Yang-Mills equations, which can be rewritten
in the form
\begin{equation}
\left(\mu S\omega_j'\right)'=-\frac{SW_{k,j}\omega_j}{r^2}.
\end{equation}
Let $\mu\omega'_j\equiv y$. Then we can write the above equation
as
\begin{equation}
(Sy)'=-\frac{SW_{k,j}\omega_j}{r^2}.
\end{equation}
From earlier, we can say that $S(r)$ is finite for all $r\in\bar{I}$. Now we
have
\begin{equation}\label{me}
S(r_1)y(r_1)-S(r_0)y(r_0)=-\int^{r_1}_{r_0}S\frac{W_{k,j}\omega_j}{r^2}dr.
\end{equation}

We can now apply the Cauchy-Schwarz inequality again to obtain
\begin{equation}
\left(\int^{r_1}_{r_0}S\frac{W_{k,j}\omega_j}{r^2}dr\right)^2\leq\int^{r_1}_{r_0}S^2dr\int^{r_1}_{r_0}\frac{W^2_{kj}\omega^2_j}{r^4}dr.
\end{equation}
Now, both $S(r)$ and $\frac{W_{k,j}\omega_j}{r^2}$ are finite on the interval $[r_0,r_1]$, therefore the integrals on the RHS are finite, therefore the LHS is bounded. Therefore, from \eqref{me} we get $(Sy)(r_1)$ finite, therefore $\mu(r_1)S(r_1)\omega'_j(r_1)$ is finite, and since $\mu(r_1)>0$, we finally have $\omega'_j(r_1)$ is finite.\par Therefore we have proven that if $\mu>0$, then given some initial conditions (either on or outside the event
horizon, so that $\mu>0$) to begin with, we can continue to regularly integrate out to obtain a solution.$\Box$\\

We now have all the pieces we need to `patch together' solutions which exist and are regular throughout the range of the spacetime. The exact nature of this patching together will be discussed at the end. We now briefly turn our attention to the behaviour of the equations in the asymptotic limit.

\subsection{Asymptotic behaviour of solutions in adS space}

\label{asymp}

The main reason for the abundance of black hole solutions in the $\Lambda<0$ case (as opposed to the $\Lambda=0$ case) is the behaviour of the field equations in the asymptotic limit $r\rightarrow\infty$. This section closely follows our work in \cite{1}.

As $r\rightarrow\infty$, the Yang-Mills equations \eqref{FEsunkom} approximate to:
\begin{equation}\label{autfes}
r^2\left(-\frac{\Lambda r^4}{3}\right)\omega''_j-\frac{2\Lambda
r^3}{3}\omega'_j+W_{k,j}\omega_j=0.
\end{equation}
We attempt to make the equations autonomous. First we introduce a
new variable $\tau$ \cite{1} such that
\begin{equation}\label{tauvar}
\tau=\sqrt{-\frac{3}{\Lambda}}\frac{1}{r},
\end{equation}
which reduces the equations \eqref{autfes} down to
\begin{equation}
\ddot{\omega_j}+W_{k,j}\omega_j=0.
\end{equation}
This tells us that the system has a critical point when
\begin{equation}
\left(k-\omega^2_j+\frac{1}{2}\left(\omega_{j+1}^2+\omega_{j-1}^2\right)\right)\omega_j=0,\qquad
j=1\ldots N-1.
\end{equation}
This constraint is identical to \eqref{wcon}, so we already know the solutions (given by \eqref{Omsol}. For $k=1$, the solutions are

\begin{equation}
\begin{matrix}
\omega_j=0 & \mbox{ or } & \omega_j=\pm\sqrt{j(N-j)}.
\end{matrix}
\end{equation}

For $k\neq 1$ in general $\mathfrak{su}(N)$, the only critical point is at $\omega_j=0$. However, as in previous works (e.g. \cite{1, 12}), it is not the nature of the critical points that is important, so much as the choice of variable \eqref{tauvar} we used to make the equations autonomous. In the $\Lambda=0$ case, an appropriate choice of radial parameter was $\tau\propto-\log r$ (see e.g. \cite{6}), which goes to infinity as $r$ goes to infinity, so that the system had to reach the critical point (as it had to go `all the way along' its trajectory in the phase space.) In our case, however, we use $\tau\propto1/r$ as being more appropriate. As our radial
parameter $r$ goes to infinity, the trajectory on the phase space gets shorter and shorter, so it does not have to reach the critical point of the system at infinity. To put it another way, a solution which corresponds to one within some interval $r\in [r_1,\infty)$ (for some large value of $r_1$) will be transformed to one in the interval $(0, \tau_1]$, and $\tau_1$ is arbitrarily small for large $r_1$. This also explains why the values of the $\omega_j$s are allowed to be arbitrary at infinity, rather than being constrained to continue along a trajectory until reaching a limit defined by the position of the critical points.

\subsection{Existence of non-trivial $\mathfrak{su}(N)$ solutions}

Now have all of the pieces we need (Propositions \ref{lexeh} and \ref{lexinf} and Lemma \ref{globreg}) to establish the existence of non-trivial $\mathfrak{su}(N)$ solutions to the 4D topological EYM field equations. 

The first theorem we shall prove (Theorem \ref{fixrhhighL}) is one of the two main results of our paper: it establishes the existence of nodeless topological $\mathfrak{su}(N)$ solutions in the regime $|\Lambda|\rightarrow\infty$, for any fixed $r_h$ (respecting possible minimum bounds on $r_h$ if $k=-1$) and $\omega_{j,h}$. Note that Theorem \ref{fixrhhighL} is very much in line with the similar proof given in \cite{1} for $k=1$.

The next proposition proves the existence of nodeless topological $\mathfrak{su}(N)$ solutions in some sufficiently small neighbourhood of an existing such solution -- i.e. that these solutions exist in open sets -- for any $\Lambda<0$. Finally, we use this proposition and the existence of trivial solutions to the field equations to prove the existence of non-trivial, nodeless topological $\sun$ solutions for $\Lambda<0$. 

\subsection{Existence of nodeless topological $\mathfrak{su}(N)$ solutions for sufficiently large $|\Lambda|$}
\label{sunlam}

For $k=1$ it was demonstrated numerically in \cite{11} that for any value of $N$, if $|\Lambda|$ was large enough, all of the solutions were such that the gauge field functions were nodeless; and in \cite{2} it was shown for $\mathfrak{su}(2)$ that if $k\neq 1$, the gauge field is monotonically increasing and therefore nodeless, for a positive $\omega_h$. However, it may be observed, as in \cite{1}, that as $N$ increases (for fixed $\Lambda$), the region in which solutions may be found grows ever smaller. We now show that for any $N$, for fixed initial parameters at the event horizon, and for $|\Lambda|\rightarrow\infty$, corresponding solutions to the field equations exist with gauge field functions non-zero everywhere. 

It can be noted that the situation here is very similar indeed to both the $\mathfrak{su}(2)$ \cite{12} and $\mathfrak{su}(N)$ \cite{1} spherically symmetric cases for $\Lambda<0$, for which the analogue of the following proposition has already been proven; so we will only sketch out the derivation here.

\begin{thr}\label{fixrhhighL}
For fixed $r_h$ and $\omega_{j,h}$, and for $|\Lambda|$ sufficiently large, there exists a black hole solution of the $\mathfrak{su}(N)$ EYM field equations such that all the gauge field functions $\omega_j(r)$ have no zeroes.
\end{thr}
\textbf{Proof} First, a qualification: if $k = -1$, and for $\Lambda<0$, we find that there is a minimum bound on $r_h$ given by $m(r)$ being monotonic, i.e.

\begin{equation}
r_h>\sqrt{\frac{2m'_h+1}{-\Lambda}},
\end{equation}
with the RHS manifestly positive. However, it should be noted that if $|\Lambda|\rightarrow\infty$, this minimum radius $r_h\rightarrow 0$. 

We note that for fixed $r_h$ and $\omega_{j,h}$, the constraint \eqref{2m'rhk} for a regular event horizon is satisfied for all sufficiently large $|\Lambda|$. As in \cite{12, 1}, it is helpful to define a length scale $\ell$ such that $\ell^2=-3/\Lambda$, and then new variables $\tilde{m}$ and $\tilde{\mu}$, which will be finite as $|\Lambda|\rightarrow\infty$, $\ell\rightarrow 0$:

\begin{equation}\label{inftyredef}
\tilde{m}=m\ell^2,\qquad\tilde{\mu}=k\ell^2-\frac{2\tilde{m}}{r}+r^2.
\end{equation}
The field equations then take the form
\begin{equation}\label{redefFEs}
\begin{split}
\tilde{m}'&=\left(k\ell^2-\frac{2\tilde{m}}{r}+r^2\right)G+\ell^2r^2p_{\theta,k},\\
0&=r^2\left(k\ell^2-\frac{2\tilde{m}}{r}+r^2\right)\omega''_j+\left[2\tilde{m}-2\ell^2r^3p_{\theta,k}+2r^3\right]\omega'_j+\ell^2W_{k,j}\omega_j.\\
\end{split}
\end{equation}
In the limit $\ell\rightarrow 0$, these equations simplify considerably and have the unique solution
\begin{equation}
\tilde{m}(r)=\tilde{m}(r_h)=\frac{1}{2}r_h^3,\qquad S(r)=1,\qquad \omega_j(r)=\omega_j(r_h).
\end{equation}
However we would also like to extend these results to the case of $\ell$ arbitrarily small (and not just $\ell=0$) by analyticity. We continue along the lines of \cite{1}. We can use the change of variables
\begin{equation}
\tilde{\lambda}=\ell^2\lambda;
\end{equation}
then the equations \eqref{rhlocalex} are unchanged except for $x\frac{d\ell}{dx}=0$ instead of $x\frac{d\Lambda}{dx}=0$, and for $x\frac{d\tilde{\lambda}}{dx}$, we get:
\begin{equation}
x\frac{d\tilde{\lambda}}{dx}=-\tilde{\lambda}+xH_{\tilde{\lambda}}+F_{\tilde{\lambda}},
\end{equation}
where
\begin{equation}
H_{\tilde{\lambda}}=-\frac{\tilde{\lambda}}{\rho}(1+2G),\qquad F_{\tilde{\lambda}}=\frac{k\ell^2}{\rho}+3\rho-\frac{2\ell^2}{\rho^3}\mathcal{P}.
\end{equation}
We can note by looking at equations \eqref{inftyredef} and \eqref{redefFEs} that the only appearance of the constant $k$ is as a `coefficient' of terms in $\ell^2$. Therefore as $\ell\rightarrow 0$ we expect the influence of $k$ to be removed completely from the analysis, and the results are then the same as in the spherically symmetric case: the field equations are all regular as $\ell\rightarrow 0$, and the solutions are analytic in $r_h$, $\omega_{j,h}$ and $\ell$.

Near $r=\infty$, exactly the same argument used to attain Proposition \ref{lexinf} shows we have solutions for $\ell$ small, and such solutions will be analytic in the field variables. Therefore, we choose some values for $r_h$ and $\omega_{j,h}$ and fix some $r_1 \gg r_h$. When we vary $\ell$ we see that solutions exist for arbitrarily small $\ell$ for which all of the gauge field functions have no zeroes in the interval $[r_h,r_1]$. Finally if $r_1$ is large enough we can use the previously discussed fact that in the asymptotic region we must use a parameter like $\tau \propto 1/r$ to show that as $r\rightarrow\infty$, the solutions will remain regular and nodeless.$\Box$

\subsection{Existence of nodeless topological $\mathfrak{su}(N)$ solutions for fixed $\Lambda<0$ in a neighbourhood of trivial solutions}\label{exsolneightriv}

We now show that solutions exist in another regime. First, we prove a powerful proposition: that if we assume the existence of an $\mathfrak{su}(N)$ black hole solution, then we can definitely find other such solutions for any $\Lambda<0$ in some sufficiently small neighbourhood of it, which possess the same number of nodes. That is, $\mathfrak{su}(N)$ solutions exist in open sets which all have the same number of nodes.

\begin{prop}\label{sunneighb}
Assume we have an existing topological $\mathfrak{su}(N)$ solution of the field equations, with the gauge field functions all nodeless for $k\neq 1$, and the initial gauge field values given by $\omega_{1,h}$, $\omega_{2,h}$, ... , $\omega_{N-1,h}$. Then all initial gauge field values in a neighbourhood of these values will give an $\mathfrak{su}(N)$ solution to the field equations in which all the gauge field functions are nodeless.
\end{prop}     

\textbf{Proof} Assume we know of a non-trivial nodeless topological $\mathfrak{su}(N)$ black hole solution of the field equations with event horizon radius $r_h$ and initial gauge field values at the event horizon $\omega_{j,h}$. As we pointed out in Section \ref{secbcs} we are only interested in black holes as solitons can be shown not to exist for $k\neq 1$. So using these initial conditions and the field equations (\ref{FEsunkom}, \ref{FEsunkmpr}, \ref{FEsunkd}), we can integrate out and get a solution that is regular all the way to infinity. For the rest of this argument we assume that $|\Lambda|$ and $r_h$ are fixed and that the gauge field functions $\omega_j$ are all nodeless. 

From the local existence theorems we proved (Propositions \ref{exrh}, \ref{lexinf}), we know that for any set of initial values of the gauge field $\omega_{1,h}$, ..., $\omega_{N-1,h}$, there are solutions locally near the event horizon, and that the solutions are analytic in their choice of initial values. For the existing $\mathfrak{su}(N)$ solution, it must be true that $\mu(r)>0$ on the interval $[r_h,\infty)$. So, by analyticity of solutions, the nearby $\mathfrak{su}(N)$ solution with initial gauge field values close to those of the existing solution, will all have $\mu(r)>0$ for all $r\in[r_h,r_c]$ for some $r_c$. By Lemma \ref{globreg}, these nearby solutions will also be regular on $[r_h,r_c]$.

Now take some $r_1\gg r_h$ such that for the existing solution, $m(r_1)$ satisfies $m(r_1)/r_1\ll 1$; and let $\tilde{\omega}_{j,h}$ be a different set of initial gauge field values (for another $\sun$ solution) in some sufficiently small neighbourhood of the $\omega_{j,h}$s. By the analyticity argument above, these nearby $\mathfrak{su}(N)$ solutions evolved from $\tilde{\omega}_{j,h}$ will be regular in the interval $[r_h,r_1]$. That is, in this interval, it is again true that $\mu>0$ and all the gauge field functions are nodeless.

Furthermore, it will still be the case at $r_1$ that $m(r_1)/r_1\ll 1$ for these nearby $\sun$ solutions as well as the existing $\sun$ solution. Because of this, and because $r_1\gg r_h$, we are able to consider the solutions in the asymptotic regime. Provided $r_1$ is sufficiently large, the solution (parameterised proportional to $1/r$) will not move very far along the phase space trajectory as $r$ increases from $r_1$ towards infinity. Therefore $m(r)/r$ will continue to be small and the asymptotic regime will continue to be valid.$\Box$\\

Lastly, we will use Proposition \ref{sunneighb} to prove the second of our main results -- the existence of genuinely non-trivial $\mathfrak{su}(N)$ solutions, for any fixed $\Lambda<0$, in some neighbourhood of certain trivial solutions (see Section \ref{secembed}). 

\begin{thr}\label{neighproof}
There exist non-trivial black hole solutions to the $\sun$ topological EYM equations (\ref{FEsunkom}, \ref{FEsunkmpr}, \ref{FEsunkd}), regular throughout the range $r_h<r<\infty$. These solutions are analytic in the parameters $r_h$, $\omega_{j,h}$ and $m(r_h)$, and are nodeless.
\end{thr}

\textbf{Proof} From Section \ref{secembed}, it is clear that there exist RNadS-type trivial solutions to the topological $\essu$ equations. From Proposition \ref{sunneighb}, we can therefore infer the existence of nearby $\essu$ solutions which are non-trivial (i.e. whose gauge field functions are not identically zero). Furthermore, from van der Bij and Radu \cite{2} we see that that with $\essu$ black holes for $k\neq 1$, $\omega$ (which we can assume without loss of generality is positive at the event horizon) is monotonically increasing, so it can never equal zero; hence all these $\essu$ solutions will be nodeless. 

Therefore, we are also able to take one such nodeless $\essu$ solution and embed it as a topological $\sun$ solution (remembering that this solution will similarly be nodeless, due to to \eqref{embedscale}). Finally, using Proposition \ref{sunneighb} once again, we may infer the existence of genuinely non-trivial and nodeless $\sun$ solutions in some sufficiently small neighbourhood of this embedded nodeless $\sun$ solution.$\Box$ 

\section{Conclusion}

The aim of this paper was to prove the existence of genuine (i.e. non-trivial and non-embedded) solutions of the 4D $\mathfrak{su}(N)$ EYM topological field equations. We began by justifying the topological metric ansatz we used. Then, we described the field equations and the gauge potential ansatz, taking a detour to prove its validity (i.e. that it is invariant under an action of $\mathfrak{su}(N)$ by principal bundle automorphisms), adapting a theorem of K\"{u}nzle's \cite{21}. We then considered the boundary conditions at the important singular points: the event horizon ($r=r_h$) and asymptotically ($r\rightarrow\infty$). We note that we considered only purely magnetic gauge fields, for which we had $N-1$ gauge field functions $\omega_j$. We also derived some trivial solutions to these equations, and described how any $\mathfrak{su}(2)$ solution can be embedded in $\mathfrak{su}(N)$ to yield a solution by a scaling of variables.  

The field equations in question are singular at the event horizon and asymptotically, thus we began by proving local existence of solutions in the neighbourhood of these two singular regimes of the equations, using a method following \cite{1, 6, 24, 12}. We also proved a lemma which states that any solution which regular in some small interval can continue to be integrated using the field equations to produce a solution which remains regular arbitrarily far from the event horizon, as long as $\mu(r)$ remains positive. We then used these to prove one of our two main results: that for fixed values of the field variables at the event horizon ($r_h$, $\omega_{j,h}$, $m_h$), then as $|\Lambda|\rightarrow\infty$ we can demonstrate the existence of solutions to the $\sun$ EYM field equations.

Next, we proved a proposition stating that given the existence of any nodeless topological $\sun$ solution, we can always find solutions arbitrarily close that possess the same number of nodes. Finally, we used the existence of known trivial solutions, and this proposition, to demonstrate the existence of non-trivial nodeless solutions to the field equations.

Our main two results, therefore, prove the existence (in the stated regimes) of four-dimensional, topological solutions to $\mathfrak{su}(N)$ EYM field equations with a negative cosmological constant, with $N-1$ gauge degrees of freedom as in the spherically symmetric case ($k=1$), previously examined in \cite{1}.

There are several very productive directions in which this research could next be taken. The most obvious and pressing issue is that of the linear stability (under non-spherically symmetric perturbations, for £$k\neq 1$) of the solutions we found: as we noted, it has previously been shown in \cite{15, 45, 46} that the number of unstable modes of a solution is related to the number of nodes the gauge fields possess, and we have proven the existence (in certain regimes) of nodeless black hole solutions.

Given that we were successfully able to generalise previous work (for black holes in $\mathfrak{su}(N)$ and $\Lambda<0$) to include spaces whose topologies give the symmetries other than spherical, it is natural to ask what other sorts of solutions we could generalise in this way; and as we said in the introduction, there is no shortage of work that we may possibly generalise. %We haven't covered the subject of alternate topologies exhaustively -- black hole and soliton solutions are known for $\mathfrak{su}(2)$ EYM theory in adS, static, axially symmetric manifolds \cite{40, 41} -- perhaps this could be extended to $\mathfrak{su}(N)$. 
We mentioned the work of Mann on topological higher-dimensional black holes \cite{23} -- again, perhaps this work could be extended to $\mathfrak{su}(N)$. 
%As noted in \cite{1}, rotating $\mathfrak{su}(2)$ EYM black holes in AdS are known to exist \cite{42}, it is possible that generalisations to topological spacetimes and to $\mathfrak{su}(N)$ exist. An area that the authors are particularly keen to explore is the possibility of generalising to even more complicated compact gauge groups -- possibly even the exceptional Lie groups -- though the challenges to finding numerical solutions in these cases would be daunting. 
%It could be rewarding to revisit the work of Oliynyk and K\"{u}nzle on the existence of solutions for general gauge groups (e.g. \cite{37}), and consider the effect of introducing $\Lambda<0$ and $k\neq 1$.% 
%Also, we note that though we have restricted our attention to purely magnetic solutions, it is possible that topological `dyonic' solutions exist -- after all, in $\mathfrak{su}(2)$ AdS, black hole and soliton solutions exist \cite{13, 14} (though their stability remains an open question) and solutions also exist for $\mathfrak{su}(N)$ in asymptotically flat space, so finding solutions with an electric part as well in $\mathfrak{su}(N)$ AdS would not be that surprising.
Finally, as in \cite{1}, since there is no theoretical upper limit on $N$, neither is there on the amount of `hair' we may give a topological black hole, though an interesting area of exploration might be $\mathfrak{su}(\infty)$: there is some evidence for existence of solutions to these \cite{26} but more work is required, and this work could extend into the inclusion of alternate topologies. Results from this may or may not be significant, especially in view of the adS/CFT (conformal field theory) correspondence (see \cite{43}, among others) and the fact that it has been conjectured that there are observables in the dual CFT which are sensitive to the presence of black hole hair (see \cite{44} for a discussion of non-Abelian solutions in the context of the adS/CFT correspondence.)

%\subsection*{Acknowledgements}

%\begin{figure}
 % \includegraphics{example.eps}
%\end{figure}
%\begin{figure*}
 % \includegraphics[width=0.75\textwidth]{example.eps}
%\end{figure*}
%\begin{table}
%\begin{tabular}{lll}
%\hline\noalign{\smallskip}
%\end{tabular}
%\end{table}

\begin{acknowledgements}
The author would like to thank Prof. E. Winstanley for many useful conversations and moral support; and Prof. H. P. K\"{u}nzle for a very useful email exchange.
\end{acknowledgements}


%\bibliography{IOP_style_topological_existence_GRG2}   % name your BibTeX data base


\end{document}